\documentclass[aps,prd,superscriptaddress,twoside,twocolumn,nofootinbib,showpacs,floatfix]{revtex4-1}
\usepackage{amsmath,amssymb}
\usepackage{graphicx,bm,braket}
\usepackage{subfigure}
\allowdisplaybreaks
\pdfpagewidth=\paperwidth
\pdfpageheight=\paperheight
\allowdisplaybreaks
\newcommand*{\fs}[1]{{#1\!\!\!/}}
\newcommand*{\hc}{\text{H.\,c.}}

\begin{document}

\title{\boldmath Analysis of the differential cross section and photon beam asymmetry data for $\gamma p \to \eta^\prime p$}

\author{Yu Zhang}
\affiliation{School of Nuclear Science and Technology, University of Chinese Academy of Sciences, Beijing 101408, China}
\author{Ai-Chao Wang}
\affiliation{School of Nuclear Science and Technology, University of Chinese Academy of Sciences, Beijing 101408, China}
\author{Neng-Chang Wei}
\affiliation{School of Nuclear Science and Technology, University of Chinese Academy of Sciences, Beijing 101408, China}
\author{Fei Huang}
\email[Corresponding author. Email: ]{huangfei@ucas.ac.cn}
\affiliation{School of Nuclear Science and Technology, University of Chinese Academy of Sciences, Beijing 101408, China}

\date{\today}

\begin{abstract}
The photoproduction reaction of $\gamma p \to \eta^\prime p$ is investigated based on an effective Lagrangian approach in the tree-level approximation, with the purpose being to understand the reaction mechanisms and to extract the resonance contents and the associated resonance parameters in this reaction. Apart from the $t$-channel $\rho$ and $\omega$ exchanges, $s$- and $u$-channel nucleon exchanges, and generalized contact term, the exchanges of a minimum number of nucleon resonances in the $s$ channel are taken into account in constructing the reaction amplitudes to describe the experimental data. It is found that a satisfactory description of the available data on both differential cross sections and photon beam asymmetries can be obtained by including in the $s$ channel the exchanges of the $N(1875)3/2^-$ and $N(2040)3/2^+$ resonances. The reaction mechanisms of $\gamma p \to \eta^\prime p$ are discussed and a prediction for the target nucleon asymmetries is presented.
\end{abstract}


\maketitle


\section{Introduction} \label{sec:Intro}

The study of nucleon resonances ($N^\ast$s) has been of great interest in hadron physics, since a deeper understanding of $N^\ast$s can help us get insight into the nonperturbative regime of quantum chromodynamics. Apart from the $\pi N$ scattering or $\pi $ photoproduction reactions where we get most of knowledge about $N^\ast$s, the production processes of heavier mesons such as $\eta$, $\eta^\prime$, $K$, $K^\ast$, $\omega$, and $\phi$ have been gaining increasing attention, especially in study of the resonances that couple weakly to $\pi N$ but strongly to other meson-baryon channels. In the present work, we concentrate on the $\eta^\prime$ photoproduction reaction. As $\eta^\prime N$ has a much higher threshold than $\pi N$, this reaction is more suitable than pion production reactions to investigate the $N^\ast$s in the less-explored higher-energy region. Furthermore, the $\eta^\prime$ photoproduction acts as an ``isospin filter," isolating the $N^\ast$s with isospin $I=1/2$.

Experimentally, increasing amounts of data on differential cross sections and photon beam asymmetries for $\gamma p \to \eta^\prime p$ became available in recent years. In 2009, the CLAS Collaboration at the Thomas Jefferson National Accelerator Facility reported the high-precision differential cross section data in the center-of-mass energy range $W\approx 1.925 \sim 2.795$ GeV \cite{CLAS09}, and the CBELSA/TAPS Collaboration released the differential cross section data in the energy range $W\approx 1.934\sim 2.35$ GeV \cite{CBELSA09}. In 2017, the A2 Collaboration at MAMI reported the differential cross section data from threshold, $W \approx 1.896$ GeV, up to $W = 1.956$ GeV \cite{MAMI}. The first photon beam asymmetry data at two center-of-mass energies, 1.903 and 1.912 GeV, were published by the GRALL Collaboration in 2015 \cite{beam2015}. Later in 2017, the CLAS Collaboration reported the photon beam asymmetry data for this reaction in a much wider energy range, $W\approx 1.9\sim 2.1$ GeV \cite{beam2017}. In 2019, the GlueX Collaboration reported the photon beam asymmetry data for this reaction at $W\approx 4.172$ GeV \cite{GlueX2019}. These photon beam asymmetry data provide much stronger constraints than the differential cross section data on constructing the reaction amplitudes for $\gamma p \to \eta^\prime p$. 

Theoretically, before the availability of the photon beam asymmetry data \cite{beam2015,beam2017,GlueX2019}, there were several works that aimed to describe the differential cross section data for $\gamma p \to \eta^\prime p$, e.g., Refs.~\cite{NH06,Zhong:2011,Huang:2013}. Nevertheless, none of them predicted correctly the experimentally observed behavior of the photon beam asymmetries for this reaction. Very recently, the photon beam asymmetry data from the GRALL and the CLAS Collaborations together with the differential cross section data for $\gamma p \to \eta^\prime p$ have been simultaneously described in partial-wave analysis performed by the BnGa group \cite{Anisovich2017,Anisovich2018} and in the updated $\eta$MAID model \cite{etaMAID2018}. The partial-wave analysis of BnGa group concluded that the contributions from the $N(1895)1/2^-$, $N(2100)1/2^+$, $N(2120)3/2^-$, and $N(1900)3/2^+$ resonances are significant in the $\eta^{\prime} p$ photoproduction reaction \cite{Anisovich2017,Anisovich2018}, while in the updated $\eta$MAID model it was claimed that the $N(1895)1/2^-$, $N(1880)1/2^+$, $N(1860)5/2^+$, and $N(1990)7/2^+$ resonances are of significance in the $\gamma p \to \eta^\prime p$ photoproduction process \cite{etaMAID2018}.

In the present work, we investigate the $\gamma p \to \eta^\prime p$ reaction within an effective Lagrangian approach at the tree-level approximation. In addition to the $t$-channel $\rho$ and $\omega$ exchanges, $s$- and $u$-channel nucleon ($N$) exchanges, and generalized contact current, we consider as few as possible $N$ resonances in the $s$ channel in constructing the reaction amplitudes to describe the available differential cross section data and photon beam asymmetry data from the CLAS, A2, and GRALL Collaborations \cite{CLAS09,MAMI,beam2015,beam2017}. The differential cross section data from the CBELSA/TAPS Collaboration \cite{CBELSA09} are not included in the fit as these data cover a narrower energy range and have less precision than the data from the CLAS Collaboration \cite{CLAS09}. The beam asymmetry data from the GlueX Collaboration \cite{GlueX2019} are not included in the present work, either. These data are measured at $W \approx 4.172$ GeV. In such a high energy region, a pure Regge model \cite{JPAC2017} might be needed rather than a Feynman model as employed in the present work.

The present paper is organized as follows. In Sec.~\ref{Sec:formalism}, we briefly introduce the framework of our theoretical model, including the treatment of gauge invariance, the effective interaction Lagrangians, the resonance propagators, and the phenomenological form factors employed in the present work. In Sec.~\ref{Sec:results}, we present our theoretical results of the differential cross sections and the photon beam asymmetries, where discussions about the reaction mechanisms are also made and a prediction of the target asymmetries for $\gamma p \to \eta^\prime p$ is given. Finally, we give a brief summary in Sec.~\ref{sec:summary}.

\section{Formalism}  \label{Sec:formalism}

\begin{figure}
\subfigure[~$s$ channel]{\label{fig:subfig:a} \includegraphics[width=0.45\linewidth]{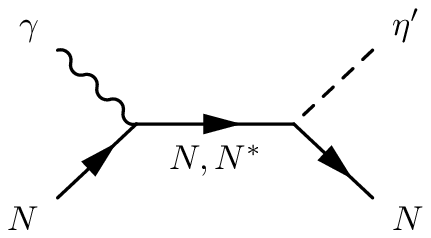}} 	\hspace{0.01\linewidth}
\subfigure[~$t$ channel]{\label{fig:subfig:b} \includegraphics[width=0.45\linewidth]{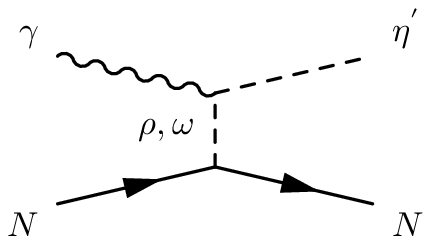}} 	\vfill
\subfigure[~$u$ channel]{\label{fig:subfig:c} \includegraphics[width=0.45\linewidth]{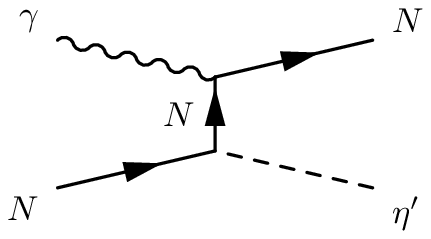}} 	\hspace{0.01\linewidth}
\subfigure[~Interaction current]{\label{fig:subfig:d} \includegraphics[width=0.45\linewidth]{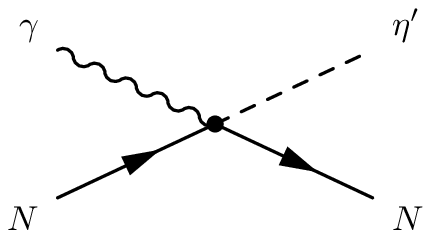}}
\caption{Generic structure of the amplitude for $\gamma N \rightarrow \eta^{\prime} N$. Time proceeds from left to right. }  \label{FIG:feymans}
\end{figure}

Following a field theoretical approach of Refs.~\cite{Haberzettl:1997,Haberzettl:2006}, the full photoproduction amplitudes for $\gamma N \rightarrow \eta^{\prime} N$ can be expressed as
\begin{equation}
M^{\mu}=M^{\mu}_s+M^{\mu}_t+M^{\mu}_u+M^{\mu}_{\rm int}. \label{eq:amplitude}
\end{equation}
Here the first three terms $M^{\mu}_s$, $M^{\mu}_t$, and $M^{\mu}_u$ stand for the $s$-, $t$-, and $u$-channel pole diagrams, respectively, with $s$, $t$, and $u$ being the Mandelstam variables of the internally exchanged particles. They arise from the photon attaching to the external particles in the underlying $NN\eta^{\prime}$ interaction vertex. The last term, $M^{\mu}_{\rm int}$, stands for the interaction current that arises from the photon attaching to the internal structure of the $NN\eta^{\prime}$ interaction vertex. All the four terms in Eq.~(\ref{eq:amplitude}) are diagrammatically depicted in Fig.~\ref{FIG:feymans}.

In the present work, the following contributions, as shown in Fig.~\ref{FIG:feymans}, are considered in constructing the $s$-, $t$-, and $u$-channel amplitudes: (i) $N$ and $N^\ast$ exchanges in the $s$ channel, (ii) $\rho$ and $\omega$ exchanges in the $t$ channel, and (iii) $N$ exchange in the $u$ channel. Using an effective Lagrangian approach, one can, in principle, obtain explicit expressions for these amplitudes by evaluating the corresponding Feynman diagrams. However, the exact calculation of the interaction current $M^{\mu}_{\rm int}$ is impractical, as it obeys a highly nonlinear equation and contains diagrams with very complicated interaction dynamics. Following Refs.~\cite{Haberzettl:2006,Huang:2012,Huang:2013}, we model the interaction current by a generalized contact current, 
\begin{equation}
M^{\mu}_{\rm int}=\Gamma_{NN\eta^{\prime}} C^\mu.   \label{eq:GCC}
\end{equation}
Here $\mu$ is the Lorentz index for $\gamma$, and $\Gamma_{NN\eta^{\prime}}$ is the vertex function of $N N \eta^{\prime}$ coupling given by the Lagrangian of Eq.~(\ref{eq:lambda2}),
 \begin{equation}
\Gamma_{NN\eta^{\prime}}=g_{NN\eta^{\prime}} \gamma_5.
\end{equation} 
$C^\mu$ is an auxiliary current, which is nonsingular, introduced to ensure that the full photoproduction amplitudes of Eq.~(\ref{eq:amplitude}) are fully gauge invariant. Following Refs.~\cite{Haberzettl:2006,Huang:2012}, we choose $C^\mu$ for $\gamma N \rightarrow \eta^{\prime} N$ as
 \begin{equation}
C^\mu=-Q_N\frac{f_u-\hat{F}}{u-p'^2}(2p'-k)^\mu-Q_N\frac{f_s-\hat{F}}{s-p^2}(2p+k)^\mu,
\end{equation} with
\begin{equation}
\hat{F}=1-\hat{h}(1-f_s)(1-f_u).
\end{equation}
Here $p$, $p'$, and $k$ are four momenta of the incoming $N$, outgoing $N$, and incoming photon, respectively; $Q_N$ is the electric charge of $N$; $f_s$ and $f_u$ are the phenomenological form factors for the $s$-channel $N$ exchange and $u$-channel $N$ exchange, respectively; $\hat{h}$ is an arbitrary function of $s$, $u$, and $t$, and it goes to unity in high-energy limit to prevent the violation of scaling \cite{sd1972}. In the present work, as usual \cite{Haberzettl:2006,Huang:2012}, we treat $\hat{h}$ as a fit parameter. It was found that the fitted value of $\hat{h}$ is very close to $1$. Thus, to reduce the number of fitting parameters, we simply fix $\hat{h}=1$ instead of treating it as a fit constant. Actually, as we shall see in Sec.~\ref{Sec:results}, the contributions from the interaction current [cf. Eq.~(\ref{eq:GCC})] are rather small in the present work.

Note that the $u$-channel $N^\ast$ exchanges are neglected in the present work. These contributions are expected to be rather small, as can be seen from the data on differential cross sections in the high-energy backward angles. We mention that neglecting the $u$-channel $N^\ast$ exchanges will not affect the gauge invariance of the full reaction amplitudes, since the transition Lagrangians for $N^\ast \to N \gamma$ are transverse [cf. Eqs.~(\ref{Eq:L_RNr_1hf})-(\ref{Eq:L_RNr_7hf})].

In the rest of this section, we present the effective Lagrangians, resonance propagators, phenomenological form factors, and Reggeized treatment of the $t$-channel $\rho$ and $\omega$ exchanges employed in the present work.

\subsection{Effective Lagrangians} \label{Sec:Lagrangians}

The effective interaction Lagrangians used in the present work are given below. For further convenience, we define the operators
\begin{equation}
\Gamma^{(+)} = \gamma_5  \quad  \text{and}  \quad  \Gamma^{(-)}=1,  \label{gamma}
\end{equation}
and the field-strength tensor
\begin{equation}
F^{\mu\nu} = \partial^{\mu}A^\nu-\partial^{\nu}A^\mu,
\end{equation}
with $A^\mu$ denoting the electromagnetic field.

The electromagnetic interaction Lagrangians required to calculate the nonresonant Feynman diagrams are
\begin{eqnarray}
\mathcal{L}_{NN\gamma} &=& - e\bar{N}\left[\left(\hat{e}\gamma^\mu-\frac{\kappa}{2M_N}\sigma^{\mu\nu}\partial_\nu\right) A_\mu\right] N \label{eq:L_rNNEt},  \\[6pt]
\mathcal{L}_{\gamma\rho\eta^{\prime}} &=& e\frac{g_{\gamma\rho\eta^{\prime}}}{M_{\eta^{\prime}}}\varepsilon_{\alpha\mu\lambda\nu}(\partial^\alpha A^\mu)(\partial^\lambda \eta^{\prime})\rho^\nu,  \\[6pt]
\mathcal{L}_{\gamma\omega\eta^{\prime}} &=&e\frac{g_{\gamma\omega\eta^{\prime}}}{M_{\eta^{\prime}}}\varepsilon_{\alpha\mu\lambda\nu}(\partial^\alpha A^\mu)(\partial^\lambda \eta^{\prime})\omega^\nu,
\end{eqnarray}
where $e$ is the elementary charge unit; $\hat{\kappa}_N = \kappa_p\left(1+\tau_3\right)/2 + \kappa_n\left(1-\tau_3\right)/2$, with the anomalous magnetic moments $\kappa_p=1.793$ and $\kappa_n=-1.913$; $M_{N}$ and $M_{\eta^{\prime}}$ stand for the masses of $N$ and $\eta^{\prime}$, respectively; $\varepsilon^{\alpha \mu \lambda \nu}$ is the totally antisymmetric Levi-Civita tensor with $\varepsilon^{0123}=1$. The meson-meson electromagnetic transition coupling constants, $g_{\gamma\rho\eta^{\prime}} = 1.25$ and $g_{\gamma\omega\eta^{\prime}} = 0.44$, are extracted from a systematic analysis of the radiative decay of pseudoscalar and vector mesons based on flavor SU(3) symmetry considerations in conjunction with vector-meson dominance arguments \cite{Huang:2013}.

The effective Lagrangians for meson-baryon interactions are
\begin{eqnarray}
\mathcal{L}_{NN\eta^{\prime}} &=& - i g_{NN\eta^{\prime}}\bar{N} \gamma_5 \eta^{\prime}N , \label{eq:lambda2} \\[6pt]
\mathcal{L}_{NN\rho} &=& - g_{NN\rho}\bar{N}{\left(\gamma^\mu-\kappa_\rho\frac{\sigma^{\mu\nu}}{2M_N}\partial_\nu\right)\rho_\mu}N , \\[6pt]
\mathcal{L}_{NN\omega} &=& - g_{NN\omega}\bar{N}{\left(\gamma^\mu-\kappa_\omega \frac{\sigma^{\mu\nu}}{2M_N} \partial_\nu \right) \omega_\mu} N,
\end{eqnarray}
where $g_{NN\eta^\prime}$ is treated as a fit parameter, and $g_{NN\omega} = 11.76$, $g_{NN\rho}=3.25$, $\kappa_\rho=6.1 $, and $\kappa_\omega=0$ are taken from Ref.~\cite{Huang:2013}. 

The resonance-nucleon-photon transition Lagrangians are
\begin{eqnarray}
{\cal L}_{RN\gamma}^{1/2\pm} &=& e\frac{g_{RN\gamma}^{(1)}}{2M_N}\bar{R} \Gamma^{(\mp)}\sigma_{\mu\nu} \left(\partial^\nu A^\mu \right) N  + \hc, \label{Eq:L_RNr_1hf}  \\[6pt]
{\cal L}_{RN\gamma}^{3/2\pm} &=& -\, ie\frac{g_{RN\gamma}^{(1)}}{2M_N}\bar{R}_\mu \gamma_\nu \Gamma^{(\pm)}F^{\mu\nu}N \nonumber \\
&& +\, e\frac{g_{RN\gamma}^{(2)}}{\left(2M_N\right)^2}\bar{R}_\mu \Gamma^{(\pm)}F^{\mu \nu}\partial_\nu N + \hc,  \label{Eq:L_RNr_3hf} \\[6pt]
{\cal L}_{RN\gamma}^{5/2\pm} &=& e\frac{g_{RN\gamma}^{(1)}}{\left(2M_N\right)^2}\bar{R}_{\mu \alpha}\gamma_\nu \Gamma^{(\mp)}\left(\partial^{\alpha} F^{\mu \nu}\right)N \nonumber \\
&& \pm\, ie\frac{g_{RN\gamma}^{(2)}}{\left(2M_N\right)^3}\bar{R}_{\mu \alpha} \Gamma^{(\mp)}\left(\partial^\alpha F^{\mu \nu}\right)\partial_\nu N \nonumber \\
&& +\, \hc,   \label{Eq:L_RNr_5hf}  \\[6pt]
{\cal L}_{RN\gamma}^{7/2\pm} &=&  i e\frac{g_{RN\gamma}^{(1)}}{\left(2M_N\right)^3}\bar{R}_{\mu \alpha \beta}\gamma_\nu \Gamma^{(\pm)}\left(\partial^{\alpha}\partial^{\beta} F^{\mu \nu}\right)N \nonumber \\
&& -\, e\frac{g_{RN\gamma}^{(2)}}{\left(2M_N\right)^4}\bar{R}_{\mu \alpha \beta} \Gamma^{(\pm)} \left(\partial^\alpha \partial^\beta F^{\mu \nu}\right) \partial_\nu N  \nonumber \\
&&  +\,   \hc,   \label{Eq:L_RNr_7hf}
\end{eqnarray}
where $R$ designates the nucleon resonance, and the superscript of ${\cal L}_{RN\gamma}$ denotes the spin and parity of the resonance $R$. The coupling constants $g_{RN\gamma}^{(i)}$ $(i=1,2)$ can, in principle, be determined by the resonance radiative decay amplitudes. Nevertheless, since the resonance hadronic coupling constants are unknown due to the lack of experimental information on the resonance decay to $N\eta^{\prime}$, we treat the products of the electromagnetic and hadronic coupling constants---which are relevant to the production amplitudes---as fit parameters in the present work.

The effective Lagrangians for resonance hadronic vertices can be written as
\begin{eqnarray}
{\cal L}^{1/2\pm}_{RN\eta^{\prime}} &=& \mp i g_{RN\eta^{\prime}} \bar{R} \Gamma^{(\pm)} \eta^{\prime} N + \, \hc,  \\[6pt]
{\cal L}_{RN\eta^{\prime}}^{3/2\pm} &=& \frac{g_{RN\eta^{\prime}}}{M_{\eta^{\prime}}} \bar{R}_\mu \Gamma^{(\mp)} (\partial^\mu\eta^{\prime}) N + \, \hc, \\[6pt]
{\cal L}_{RN\eta^{\prime}}^{5/2\pm} &=& \pm i \frac{g_{RN\eta^{\prime}}}{M_{\eta^{\prime}}^2}\bar{R}_{\mu \nu}\Gamma^{(\pm)} (\partial^\mu \partial^\nu \eta^{\prime}) N  + \, \hc, \\[6pt]
{\cal L}_{RN\eta^{\prime}}^{7/2\pm} &=& - \frac{g_{RN\eta^{\prime}}}{M_{\eta^{\prime}}^3}\bar{R}_{\mu \nu \alpha} \Gamma^{(\mp)} (\partial^\mu \partial^\nu \partial^\alpha \eta^{\prime}) N  + \, \hc,
\end{eqnarray}
where the coupling constants $g_{RN\eta^{\prime}}$, as mentioned above, are combined with the resonance electromagnetic coupling constants $g_{RN\gamma}^{(i)}$ $(i=1,2)$. The products of them are determined by a fit to the available data on differential cross sections and photon beam asymmetries for $\gamma p \to \eta^\prime p$.

\subsection{Resonance propagators}

The prescriptions of the propagators for resonances with spin-$1/2$, -$3/2$, -$5/2$, and -$7/2$ adopted in the present work are \cite{Wang:2017,Wang:2018,Wei2019,ac2020}
\begin{eqnarray}
S_{1/2}(p) &=& \frac{i}{\fs{p} - M_R + i \Gamma_R/2}, \\[6pt]
S_{3/2}(p) &=&  \frac{i}{\fs{p} - M_R + i \Gamma_R/2} \left( \tilde{g}_{\mu \nu} + \frac{1}{3} \tilde{\gamma}_\mu \tilde{\gamma}_\nu \right),  \\[6pt]
S_{5/2}(p) &=&  \frac{i}{\fs{p} - M_R + i \Gamma_R/2} \,\bigg[ \, \frac{1}{2} \big(\tilde{g}_{\mu \alpha} \tilde{g}_{\nu \beta} + \tilde{g}_{\mu \beta} \tilde{g}_{\nu \alpha} \big)  \nonumber \\
&& -\, \frac{1}{5}\tilde{g}_{\mu \nu}\tilde{g}_{\alpha \beta}  + \frac{1}{10} \big(\tilde{g}_{\mu \alpha}\tilde{\gamma}_{\nu} \tilde{\gamma}_{\beta} + \tilde{g}_{\mu \beta}\tilde{\gamma}_{\nu} \tilde{\gamma}_{\alpha}  \nonumber \\
&& +\, \tilde{g}_{\nu \alpha}\tilde{\gamma}_{\mu} \tilde{\gamma}_{\beta} +\tilde{g}_{\nu \beta}\tilde{\gamma}_{\mu} \tilde{\gamma}_{\alpha} \big) \bigg],   \\[6pt]
S_{7/2}(p) &=&  \frac{i}{\fs{p} - M_R + i \Gamma_R/2} \, \frac{1}{36}\sum_{P_{\mu} P_{\nu}} \bigg( \tilde{g}_{\mu_1 \nu_1}\tilde{g}_{\mu_2 \nu_2}\tilde{g}_{\mu_3 \nu_3} \nonumber \\ 
&& -\, \frac{3}{7}\tilde{g}_{\mu_1 \mu_2}\tilde{g}_{\nu_1 \nu_2}\tilde{g}_{\mu_3 \nu_3} + \frac{3}{7}\tilde{\gamma}_{\mu_1} \tilde{\gamma}_{\nu_1} \tilde{g}_{\mu_2 \nu_2}\tilde{g}_{\mu_3 \nu_3} \nonumber \\  
&& -\, \frac{3}{35}\tilde{\gamma}_{\mu_1} \tilde{\gamma}_{\nu_1} \tilde{g}_{\mu_2 \mu_3}\tilde{g}_{\nu_2 \nu_3} \bigg),  \label{propagator-7hf}
\end{eqnarray}
where
\begin{eqnarray}
\tilde{g}_{\mu \nu} &=& -\, g_{\mu \nu} + \frac{p_{\mu} p_{\nu}}{M_R^2}, \\[3pt]
\tilde{\gamma}_{\mu} &=& \gamma^{\nu} \tilde{g}_{\nu \mu} = -\gamma_{\mu} + \frac{p_{\mu}\fs{p}}{M_R^2}.
\end{eqnarray}
Here $M_R$ and $\Gamma_R$ are, respectively, the mass and width of resonance $R$, and $p$ is the resonance four momentum. The summation over $P_{\mu}(P_{\nu})$ in Eq.~(\ref{propagator-7hf}) goes over the $3! = 6$ possible permutations of the indices $\mu_1\mu_2\mu_3(\nu_1\nu_2\nu_3)$. 

In the present work, the energy-dependent resonance width is adopted in resonance propagators. Following Refs.~\cite{walker1969,RA1990,AI1997,DD1999}, the resonance width $\Gamma$ is written as a function of $W = \sqrt{s}$ in the form of
\begin{equation}
\Gamma(W) = \Gamma_R \left[ \sum_{i=1}^{N} \beta_i \hat{\Gamma}_i(W) + \beta_\gamma \hat{\Gamma}_{\gamma}(W) \right],
\end{equation}
where the sum over $i$ accounts for decays of the resonance into two- or three-hadron channels. The $\Gamma_R$ is the total static resonance width and the numerical factors $\beta_i$ and $\beta_\gamma$ describe the resonance branching ratios into the various decay channels,
\begin{equation}
\sum_{i=1}^{N} \beta_i + \beta_\gamma = 1.
\end{equation}
The resonance width functions $\hat{\Gamma}_i(W)$ for the decay of the resonance into hadronic fragments with masses $m_{i1}$ and $m_{i2}$ are taken as
\begin{equation}
\hat{\Gamma}_i(W) = \left( \frac{q_i}{q_{iR}} \right)^{2L+1} \left(\frac{\lambda_i^2+q_{iR}^2}{\lambda_i+q_i^2}\right)^L
\end{equation}
for $W > m_{i1} +m_{i2} $, and zero otherwise. For the decay of the resonance into one baryon and two mesons, we use
\begin{equation}
\hat{\Gamma}_i(W) = \left( \frac{q_i}{q_{iR}} \right)^{4L+2} \left(\frac{\lambda_i^2+q_{iR}^2}{\lambda_i+q_i^2} \right)^{L+2}.
\end{equation}
Here
\begin{equation}
q_i(W)= \frac{\sqrt{[W^2-(m_{i1}+m_{i2})^2][W^2-(m_{i1}-m_{i2})^2]}}{2W}
\end{equation}  
and $q_{iR}=q_i(M_R)$. The $m_{i2}$ should be understood as the sum of the two meson masses in the second case. For the resonance decay into a nucleon and a photon with three-momentum $k$,
\begin{equation}
\hat{\Gamma}_{\gamma}(W) = \left(\frac{k}{k_{R}}\right)^{2L+2} \left(\frac{\lambda_\gamma^2+k_{R}^2}{\lambda_{\gamma} + k^2}\right)^{L+1}, 
\end{equation}
with
\begin{equation}
k(W)=\frac{W^2-m_N^2}{2W},
\end{equation} 
and $k_{R}=k(M_R)$. The parameters $\lambda_i$ and $\lambda_\gamma$ are set to be 1 fm$^{-1}$ for all channels. See Refs.~\cite{NH06,Huang:2013} for more details.

\subsection{Form factors}

Each hadronic vertex obtained from the Lagrangians given in Sec.~\ref{Sec:Lagrangians} is accompanied with a phenomenological form factor to parametrize the structure of the hadrons and to normalize the behavior of the production amplitude. Following Refs.~\cite{Wang:2017,Wang:2018,Wei2019,ac2020}, for intermediate baryon exchange we take the form factor as
\begin{equation}
f_B(p^2) = \left(\frac{\Lambda_B^4}{\Lambda_B^4+\left(p^2-M_B^2\right)^2}\right)^2,  \label{eq:ff_B}
\end{equation}
where $p$ and $M_B$ denote the four momentum and the mass of the exchanged baryon $B$, respectively. The cutoff mass $\Lambda_B$ is treated as a fit parameter for each exchanged baryon, except for the $N$ exchanges in the $s$ and $u$ channels, where $\Lambda_{N} = 900$ MeV is adopted. For intermediate meson exchange, we take the form factor as
\begin{equation}
f_M(q^2) = \left(\frac{\Lambda_M^2-M_M^2}{\Lambda_M^2-q^2}\right)^2,     \label{eq:ff_M}
\end{equation}
where $q$ and $M_M$ represent the four momentum and the mass of the intermediate meson $M$, respectively. The cutoff mass $\Lambda_M$ is treated as a fit parameter for each exchanged meson. 

Note that the gauge-invariance feature of our photoproduction amplitude is independent of the specific form of the form factors.

\subsection{Reggeized $t$-channel amplitudes}  \label{subsec:4}

A Reggeization of the $t$-channel amplitudes for $\rho$ and $\omega$ exchanges, which is introduced to take into account the effects of high spin meson exchanges, corresponds to the following replacement of the Feynman propagators \cite{reggewangac}
\begin{eqnarray}
\dfrac{1}{t - m_v^2} \quad \Longrightarrow \quad  \mathcal{P}_{\rm{Regge}}^v  &=&  \left(\dfrac{s}{s_0}\right)^{\alpha _v(t) - 1} \dfrac{\pi \alpha _v^\prime}{{\rm sin}\left[\pi \alpha _v(t)\right]} \nonumber \\
&&  \times \,  \dfrac{e^{-i\pi \alpha _v(t)}}{\Gamma\left[\alpha _v(t)\right]},  \label{regge_1}
\end{eqnarray}
where $v$ is vector-meson $\rho$ or $\omega$, $s_0$ is a mass scale that is conventionally taken as $s_0 = 1$ GeV$^2$, the phase $e^{-i\pi \alpha _v(t)}$ is chosen to account for degenerate trajectories and $\alpha _v^\prime$ is the slope of the Regge trajectory $\alpha _v(t) $. For $\rho$ and $\omega$, the trajectories are parametrized as \cite{regge2003}
\begin{eqnarray}
\alpha_\rho (t) &=& 0.55 + 0.8 \ {\rm GeV^{-2}} \ t,      \\[4pt]
\alpha_\omega (t) &=& 0.44 + 0.9 \ {\rm GeV^{-2}} \ t.
\end{eqnarray}
One observes that the Reggeization of the amplitudes for $t$-channel $\rho$ and $\omega$ exchanges in Eq.~(\ref{regge_1}) is equivalent to the following replacement of the form factors in the corresponding Feynman amplitudes
\begin{eqnarray}
f_t \quad \Longrightarrow \quad \mathcal{F}_t = \left(t - m_v^2\right) \mathcal{P}_{\rm Regge}^v f_t. \label{f_t_replace1}
\end{eqnarray}
It is well known that the Regge amplitudes work properly in the very-large-$s$ and very-small-$|t|$ region, while the Feynman amplitudes work properly in the low energy region. In the present work, we employ an interpolating form factor to parametrize the smooth transition from the Feynman amplitudes to the Regge amplitudes \cite{reggewangac}, which is fulfilled by using the following replacement of the form factors, instead of Eq.~(\ref{f_t_replace1}), in the corresponding Feynman amplitudes:
\begin{equation}
f_t  \quad \Longrightarrow \quad \mathcal{F}_{R,t} = \mathcal{F}_t R +f_t \left(1-R\right),
\end{equation}
where $R = R_sR_t$ with
\begin{eqnarray}
R_s &=& \dfrac{1}{1+e^{-(s-s_R)/s_0}},     \\[4pt]
R_t &=& \dfrac{1}{1+e^{-(t+t_R)/t_0}}.
\end{eqnarray}
Here $s_R$, $s_0$, $t_R$, and $t_0$ are parameters to be determined by fitting the data.

\section{Results and discussion}   \label{Sec:results}

As mentioned in Sec.~\ref{sec:Intro}, before the photon beam asymmetry data for $\gamma p \to \eta^\prime p$ from the GRALL and CLAS Collaborations became available in 2015 and 2017 \cite{beam2015,beam2017}, the theoretical works on $\gamma p \to \eta^\prime p$ could describe only the differential cross section data. After the availability of the photon beam asymmetry data, simultaneous descriptions of the differential cross section data and the photon beam asymmetry data for $\gamma p \to \eta^\prime p$ have only been performed by the BnGa group \cite{Anisovich2017,Anisovich2018} and in the updated $\eta$MAID model \cite{etaMAID2018}. The resonance contents reported from these theoretical works are quite different.

In the present work, we analyze the available differential cross section data and the photon beam asymmetry data for $\gamma p\to \eta^{\prime}p$ from the CLAS, A2, and GRALL Collaborations \cite{CLAS09,MAMI,beam2015,beam2017} within an effective Lagrangian approach in the tree-level approximation, with the purpose being to understand the reaction mechanisms and to extract the resonance contents and their associated parameters in this reaction. Apart from the $t$-channel $\rho$ and $\omega$ exchanges, $s$- and $u$-channel $N$ exchanges, and generalized contact current, we consider a minimum number of nucleon resonances in the $s$ channel in constructing the reaction amplitudes to describe the available data. We test different number of nucleon resonances and various combinations of them. 
We treat the resonance's mass, width, and hadronic and electromagnetic coupling constants as fit parameters. Actually, only the products of the resonance electromagnetic and hadronic coupling constants are relevant to the reaction amplitudes, which are what we really fit in practice.

\begin{table}[tb]
\caption{\label{Table:para}  Model parameters except those in the interpolating form factor. The asterisks below the resonance names denote the overall status of these resonances evaluated by PDG \cite{PDG}. The quantities in square brackets below the resonance mass $M_R$ and width $\Gamma_R$ of $N(1875)3/2^-$ are the corresponding ranges quoted in PDG. The decay branching ratios (in $\%$) in bold font for $N(1875)3/2^-$ denote the centroid values of the dominant decay modes quoted by PDG. For $N(2040)3/2^+$, $\beta_{p\gamma}$ is fixed to be $0.2\%$ as there is no information from PDG. }
\renewcommand{\arraystretch}{1.2}
\begin{tabular*}{\columnwidth}{@{\extracolsep\fill}lcc}
\hline\hline
$g_{NN\eta^\prime}$  	&   $0.83 \pm 0.01$ 	&	\\
$\Lambda_\rho$  [MeV]  	&   $864 \pm 1$  	&	\\
$\Lambda_\omega$  [MeV]  	&   $517 \pm 1$  	&	\\
\hline
$N^\ast$ Name  	&   $N(1875)3/2^-$  &  $N(2040)3/2^+$   \\
                      	&   $\ast\ast\ast$    &  $\ast$	\\
$M_R$  [MeV]  	&   $1853 \pm 1$   &  $2049\pm 1 $ 	\\
                             &   $[1850-1920]$  &  \\
$\Gamma_R$  [MeV]  &  $392\pm 2$  & $301 \pm 4$ 	\\
                               &  $[120-250]$   &    \\
$\beta_{p\gamma}$ [$\%$]  &   ${\bf 0.013}$   &  ${\bf 0.2}$   \\
$\beta_{N\pi}$ [$\%$]            &   ${\bf 7}$   &    \\
$\beta_{N\omega}$ [$\%$]      &   ${\bf 20}$   &    \\
$\beta_{N\pi\pi}$ [$\%$]    &   ${\bf 73}$  &  $98.51\pm 0.02$   \\
$\beta_{N\eta^\prime}$ [$\%$]   &                &  $1.29\pm 0.01$    \\
$\Lambda_{R}$ [MeV] &  $723 \pm 1$ & $1034 \pm 6$  \\
$g^{(1)}_{RN\gamma}g_{RN\eta^\prime}$   &  $ 24.75 \pm 0.10  $  & $ -1.02 \pm 0.04 $   \\
$g^{(2)}_{RN\gamma}g_{RN\eta^\prime}$  &  $ -11.93 \pm 0.16  $  & $ 4.75 \pm 0.04  $  \\
\hline\hline
\end{tabular*}
\end{table}

\begin{table}[tb]
\caption{\label{Table:para2} Fitted values of the parameters in interpolating form factor, in GeV$^2$.}
\renewcommand{\arraystretch}{1.2}
\begin{tabular*}{\columnwidth}{@{\extracolsep\fill}cccc}
\hline\hline
$s_0$ & $s_R$ & $t_0$  & $t_R$ \\
$0.49 \pm 0.04 $   & $7.80 \pm 0.04$    & $0.30 \pm 0.01$ & $2.34 \pm 0.10$ \\
\hline\hline
\end{tabular*}
\end{table}

\begin{figure*}[tbp]
\includegraphics[width=0.86\textwidth]{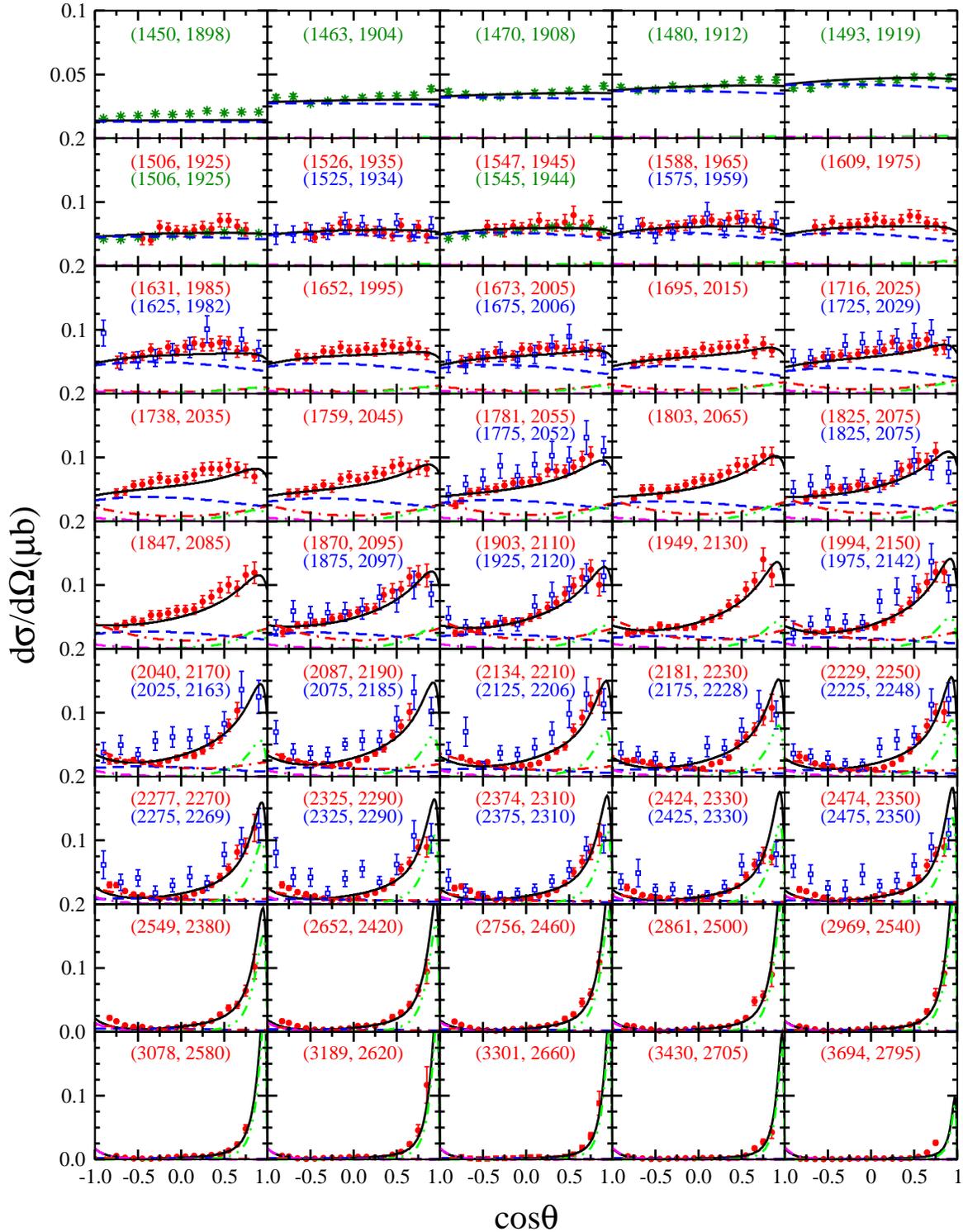}
\caption{Differential cross sections for $\gamma p\to \eta^\prime p$ as a function of $\cos\theta$. The black solid lines represent the results from the full calculation. The blue dashed, red dash-dotted, green dash-double-dotted, and magenta dot-double-dashed lines represent the individual contributions from the $s$-channel $N(1875)3/2^-$ exchange, $s$-channel $N(2040)3/2^+$ exchange, $t$-channel $\omega$ exchange, and $u$-channel $N$ exchange, respectively. The contributions from the interaction current, $t$-channel $\rho$ exchange, and $s$-channel $N$ exchange are not presented, as these contributions are too small to be clearly seen with the scale used in this figure. The green stars and red full circles denote the data from the A2 Collaboration \cite{MAMI} and CLAS Collaboration \cite{CLAS09}, respectively. The blue empty squares denote the data from the CBELSA/TAPS Collaboration \cite{CBELSA09}, which are not included in the fit. The numbers in parentheses denote the centroid values of the photon laboratory incident energy (left number) and the corresponding total center-of-mass energy of the system (right number), in MeV.}
\label{fig:dsdo_1}
\end{figure*}

\begin{figure}[tbp]
\includegraphics[width=\columnwidth]{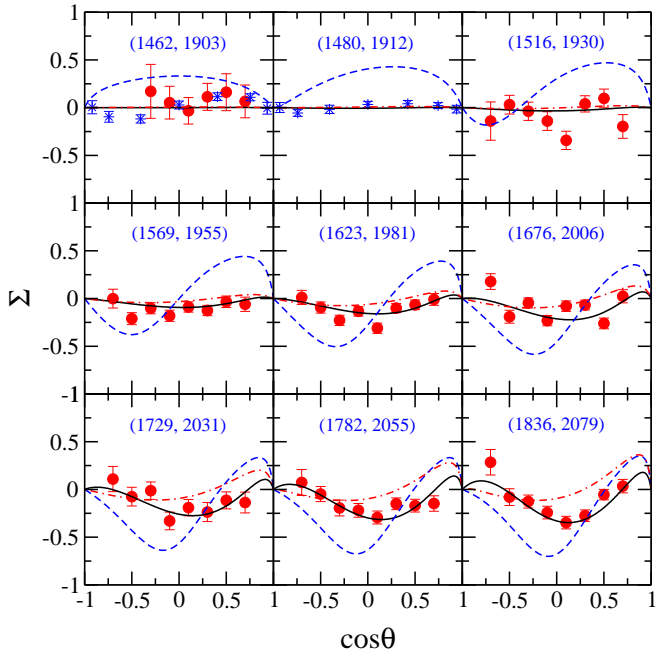}
\caption{Photon beam asymmetries as a function of $\cos\theta$ for $\gamma p\to \eta^\prime p$. The solid lines represent the results from the full amplitudes. The blue dashed and red dash-dotted lines represent the results obtained by switching off the contributions from the $N(1875)3/2^-$ and $N(2040)3/2^+$ exchanges, respectively, from the full model. The blue stars and red circles represent the data from the GRALL Collaboration \cite{beam2015} and the CLAS Collaboration \cite{beam2017}, respectively. The numbers in parentheses denote the centroid values of the photon laboratory incident energy (left number) and the corresponding total center-of-mass energy of the system (right number), in MeV.}
\label{fig:beam_1}
\end{figure}

\begin{figure}[tbp]
\includegraphics[width=\columnwidth]{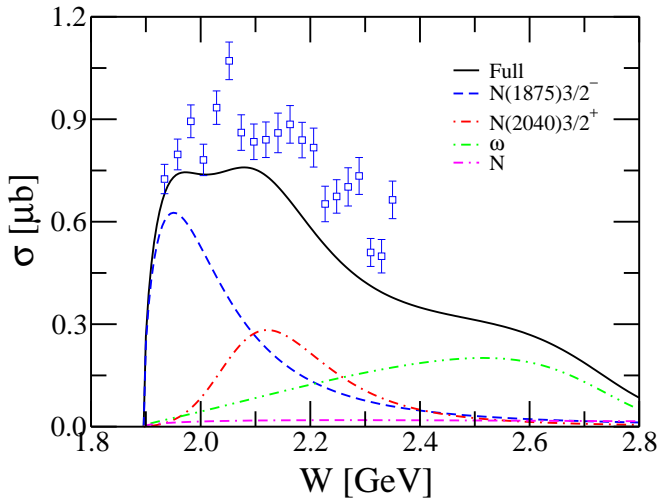}
\caption{Total cross sections with individual contributions for $\gamma p\to \eta^\prime p$. Data are from the CBELSA/TAPS Collaboration \cite{CBELSA09} but not included in the fit.}
\label{fig:sig}
\end{figure}

\begin{figure}[tbp]
\includegraphics[width=\columnwidth]{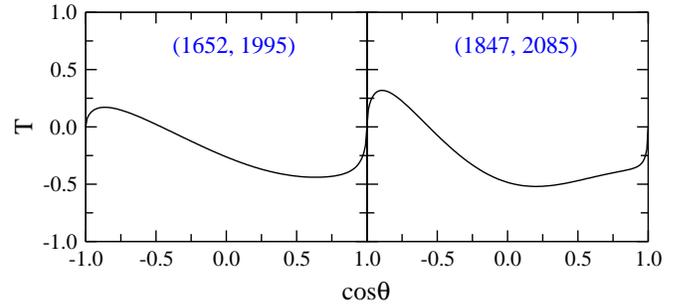}
\caption{Predicted target nucleon asymmetries as a function of $\cos\theta$ for $\gamma p\to \eta^\prime p$.}
\label{fig:tagt}
\end{figure}

We have in total $877$ data in the fit, i.e., $681$ differential cross section data from CLAS \cite{CLAS09}, $120$ differential cross section data from A2 \cite{MAMI}, $14$ photon beam asymmetry data from GRALL \cite{beam2015}, and $62$ photon beam asymmetry data from CLAS \cite{beam2017}. After numerous trials with the inclusion of different number of nucleon resonances and different combinations of them, we found that if only one resonance is taken into account in fitting the data, the $\chi^2$ per data point, $\chi^2/$N, will be larger than $4.4$ and results in a rather poor fitting quality. We then considered the exchanges of two nucleon resonances in the $s$ channel. We found that the available cross section data from the A2 and CLAS Collaborations and the beam asymmetry data from the GRALL and CLAS Collaborations can be well described by including the $N(1875){3/2}^-$ and $N(2040){3/2}^+$ resonances, which are rated as three-star and one-star resonances in PDG \cite{PDG}, respectively. The resulting $\chi^2/{\rm N}$ is $1.8$ for beam asymmetry data, $3.2$ for differential cross section data, and $3.1$ for all data. The fitted results will be shown and discussed in detail below. With the inclusion of other two resonances, the best fit will be the one with the $N(1875){3⁄2}^-$ and $N(1860){5⁄2}^+$ resonances, which results in $\chi^2/{\rm N}=2.3$ for beam asymmetry data, $4.1$ for differential cross section data, and $3.9$ for all data. It is seen that the differential cross sections at high-energy forward angles are considerably underestimated in this fit. The other fits with two resonances have even worse fitting qualities and thus are not considered as acceptable. If one more resonance apart from the $N(1875)3/2^-$ and $N(2040)3/2^+$ resonances is further taken into account, the value of $\chi^2/{\rm N}$ reduces a little bit for the differential cross section data, but it remains almost the same or gets even worse for the beam asymmetry data. Moreover, in this case, there will be many solutions with similar fitting qualities, and thus no conclusive conclusion can be drawn about the resonance contents and parameters extracted from the available limited data. Therefore, we leave the fits with three or more nucleon resonances to future work when more data on spin observables for this reaction become available. 

As discussed above, the most recent differential cross section data and photon beam asymmetry data for $\gamma p \to \eta^\prime p$ from the A2, CLAS, and GRALL Collaborations \cite{CLAS09,MAMI,beam2015,beam2017} can be satisfactorily described with the inclusion of the three-star resonance $N(1875)3/2^-$ and the one-star resonance $N(2040)3/2^+$. The model parameters determined by the fit with these two resonances are listed in Tables~\ref{Table:para} and \ref{Table:para2}. In Table~\ref{Table:para}, the asterisks below resonance names denote the overall status of these resonances evaluated by PDG \cite{PDG}. The decay branching ratios (in $\%$) in bold font are values fixed in the calculation. For the $N(1875)3/2^-$ resonance, they are fixed to the centroid values of the dominant decay modes quoted by PDG. For the $N(2040)3/2^+$ resonance, its electromagnetic branching ratio is fixed to be $0.2\%$ as there is no information from PDG for the decay of this resonance. The other adjustable parameters are determined by a fit to the available data. The quantities in square brackets below the resonance mass $M_R$ and width $\Gamma_R$ of $N(1875)3/2^-$ are the corresponding ranges quoted in PDG. One sees that our fitted mass of $N(1875)3/2^-$ is within the range of the corresponding PDG value, but the fitted width lies outside the corresponding range given by PDG. Nevertheless, our fitted width $392$ MeV is not too far away from the value $321\pm 21$ MeV given by Hunt and Manley in their partial wave analysis \cite{Hunt19}.  Note that for $s$- and $u$-channel $N$ exchange, the coupling constant $g_{NN\eta^\prime}$ is treated as a fit parameter, while the cutoff is set to be $900$ MeV in order to reduce the number of free parameters of the model. In Tables~\ref{Table:para} and \ref{Table:para2}, the uncertainties of the values of fit parameters are estimated from the uncertainties (error bars) associated with the fitted experimental differential cross section and photon beam asymmetry data. 

%

The theoretical results of the differential cross sections for $\gamma p\to \eta^\prime p$ obtained with the parameters listed in Tables~\ref{Table:para} and \ref{Table:para2} are shown in Fig.~\ref{fig:dsdo_1}. There, the black solid lines represent the results from the full calculation. The blue dashed, red dash-dotted, green dash-double-doted, and magenta dot-double-dashed lines represent the individual contributions from the $s$-channel $N(1875)3/2^-$ exchange, $s$-channel $N(2040)3/2^+$ exchange, $t$-channel $\omega$ exchange, and $u$-channel $N$ exchange, respectively. The contributions from the interaction current,  $t$-channel $\rho$ exchange, and $s$-channel $N$ exchange are not presented, as these contributions are too small to be clearly seen with the scale used in this figure. The green stars and red full circles represent the data from the A2 Collaboration \cite{MAMI} and CLAS Collaboration \cite{CLAS09}, respectively. The blue empty squares denote the data from the CBELSA/TAPS Collaboration \cite{CBELSA09} which are not included in the fit. The numbers in parentheses denote the centroid values of the photon laboratory incident energy (left number) and the corresponding total center-of-mass energy of the system (right number), in MeV. 

One sees from Fig.~\ref{fig:dsdo_1} that overall our calculated differential cross sections agree quite well with the corresponding data from the A2 and CLAS Collaborations in the entire energy region considered. Small  discrepancies are also seen in a comparison of our theoretical results with the CBELSA/TAPS data, although these data have much larger error bars, which might show that the data from the CBELSA/TAPS and CLAS Collaborations are not compatible with each other. In Ref.~\cite{Anisovich2017}, the theoretical curves were multiplied by overall scaling factors $0.9$ for differential cross sections and $0.94$ for beam asymmetries to compare with the corresponding CLAS data. In the present work, if we multiply a scaling factor $0.9$ to the CBELSA/TAPS differential cross section data, the incompatibility between the CBELSA/TAPS and CLAS datasets will become less obvious, but at high energies the CBELSA/TAPS data still tend to be larger than the CLAS data, as has already been observed in Ref.~\cite{Anisovich2017}. From Fig.~\ref{fig:dsdo_1} one also sees that, at high-energy forward angles, the differential cross sections are dominated by the $t$-channel $\omega$ exchange. The $u$-channel $N$ exchange provides small contributions at high-energy backward angles. Actually, the parameters for $t$-channel and $u$-channel interactions are determined mainly by the high-energy data at forward angles and backward angles, respectively. Then, at lower energies, one has to introduce the resonance exchanges to describe the experimental cross sections. The resonance $N(1875)3/2^-$ contributes significantly in the energy range from threshold up to $W\sim 2.3$ GeV. The resonance $N(2040)3/2^+$ provides smaller but considerable contributions in the energy range $2.0$ GeV $< W < 2.3$ GeV.

Figure~\ref{fig:beam_1} shows our theoretical results for the photon beam asymmetries compared with the data. There, the solid lines represent the results from the full amplitudes. The blue dashed and red dash-dotted lines represent the results obtained by switching off the contributions from the $N(1875)3/2^-$ and $N(2040)3/2^+$ resonances, respectively, from the full model. The blue stars and red circles represent the data from the GRALL Collaboration \cite{beam2015} and CLAS Collaboration \cite{beam2017}, respectively. One sees that our theoretical results for the photon beam asymmetries are in good agreement with the corresponding data. When the contributions from the $N(1875)3/2^-$ resonance exchange are switched off, the theoretical photon beam asymmetries deviate significantly from the data in the whole energy range considered. When the contributions from the $N(2040)3/2^+$ are switched off, considerable deviations of the theoretical beam asymmetries with the data are only seen above $W \sim 2$ GeV. This coincides with the observations from the differential cross sections as discussed above; i.e., the resonance $N(2040)3/2^+$ provides considerable contributions in the energy range $2.0$ GeV $< W < 2.3$ GeV, while the resonance $N(1875)3/2^-$ provides significant contributions in the energy range from threshold up to $W\sim 2.3$ GeV.

Figure~\ref{fig:sig} shows our predicted total cross sections (black solid lines) together with individual contributions from the $s$-channel $N(1875)3/2^-$ exchange (blue dashed line), $s$-channel $N(2040)3/2^+$ exchange (red dash-dotted line), $t$-channel $\omega$ exchange (green dash-double-dotted line), and $u$-channel $N$ exchange (magenta dot-double-dashed lines) obtained by integrating the corresponding results for differential cross sections. The contributions from the interaction current, $t$-channel $\rho$ exchange, and $s$-channel $N$ exchange are not shown, as these contributions are rather small. The data are from the CBELSA/TAPS Collaboration \cite{CBELSA09} but not included in the fit. One sees that our theoretical total cross sections are considerably lower than the data from the CBELSA/TAPS Collaboration, which is not a surprise as we have already seen from Fig.~\ref{fig:dsdo_1} that our theoretical differential cross sections do not agree well with the CBELSA/TAPS data although they are in good agreement with the data from the A2 and CLAS Collaborations. From Fig.~\ref{fig:sig} one sees that it is the $t$-channel $\omega$ exchange that dominates the background contributions of the reaction $\gamma p\to \eta^\prime p$, while the contributions from the other nonresonant terms are negligible. The resonance contributions from both $N(1875)3/2^-$ and $N(2040)3/2^+$ are significant. In particular, the $N(1875)3/2^-$ resonance dominates the sharp rise of the total cross sections near threshold and provides a bump around $W \sim 1.95$ GeV, while the $N(2040)3/2^+$ resonance is responsible for the little bump structure around $W \sim 2.1$ GeV.  

In Fig.~\ref{fig:tagt}, we show the predictions of the target nucleon asymmetries ($T$) at two selected energy points from the present model. We hope that this observable can be measured in experiments in the near future, which can help us to further constrain the theoretical model and thus to get a better understanding of the reaction mechanisms and the associated resonance contents and parameters in $\gamma p\to \eta^\prime p$.

\section{Summary and conclusion}  \label{sec:summary}

In the present work, we employ an effective Lagrangian approach at the tree-level approximation to analyze the available differential cross section and photon beam asymmetry data for the photoproduction reaction $\gamma p \to \eta^{\prime} p$. We consider the exchanges of a minimum number of nucleon resonances in the $s$ channel, in addition to the $t$-channel $\rho$ and $\omega$ exchanges, $s$- and $u$-channel $N$ exchanges, and generalized interaction current, in constructing the reaction amplitudes to describe the data. The $s$-, $u$-, and $t$-channel amplitudes are obtained by evaluating the corresponding Feynman diagrams, and the generalized contact current is constructed in such a way that the full photoproduction amplitudes are fully gauge invariant. It is found that the available differential cross section data and photon beam asymmetry data for $\gamma p \to \eta^\prime p$ from the A2, CLAS, and GRALL Collaborations \cite{CLAS09,MAMI,beam2015,beam2017} can be well reproduced by introducing the $N(1875)3/2^-$ and $N(2040)3/2^+$ resonances in the $s$-channel interactions. Further analysis shows that the $t$-channel $\omega$ exchange dominates the high-energy forward angle cross sections, and the $u$-channel $N$ exchange contributes considerably to the high-energy backward angle cross sections. The resonance $N(1875)3/2^-$ contributes significantly to both differential cross sections and photon beam asymmetries in the energy range from threshold up to $W\sim 2.3$ GeV. The resonance $N(2040)3/2^+$ contributes considerably in the energy range $2.0$ GeV $< W < 2.2$ GeV to both differential cross sections and photon beam asymmetries. The contributions from the interaction current, $s$-channel $N$ exchange, and $t$-channel $\rho$ exchange are found to be negligible. The predictions of the target nucleon asymmetry for $\gamma p \to \eta^\prime p$ from the present model are also presented. Experimental information on this observable and on other spin observables are called for to further constrain the theoretical model and thus to get a better understanding of the reaction mechanisms for this reaction. In this respect, the data for electroproduction reactions from CLAS12 in the relevant range of $W$ and up to $Q^2$ of $12$ GeV$^2$ will hopefully provide an opportunity for better extractions of resonance parameters and couplings \cite{Carman20}.

\begin{acknowledgments}
The author Y.Z. is grateful to P. Levi Sandri for his kind help with the experimental data. This work is partially supported by the National Natural Science Foundation of China under Grants No.~11475181 and No.~11635009, the Fundamental Research Funds for the Central Universities, and the Key Research Program of Frontier Sciences of Chinese Academy of Sciences under Grant No.~Y7292610K1.
\end{acknowledgments}


\end{document}